\documentstyle[12pt,openbib]{article}
\def\baselinestretch{1.}
\title{\bf Application of Renormalization to  Potential Scattering}

\author{Carlos F. de Araujo, Jr.$^a$, Lauro Tomio$^a$,\\
Sadhan K. Adhikari$^a$\thanks{John Simon Guggenheim Memorial Foundation
Fellow}, and T. Frederico$^b$\\ \\
\small $^a$ Instituto de F\'\i sica Te\'orica, 
Universidade Estadual Paulista, \\
\small 01405-900 S\~{a}o Paulo, S\~ao Paulo, Brasil \\
\small $^b$ Departamento de F\'\i sica, Instituto Tecnol\'ogico da 
Aeron\'autica, Centro T\'ecnico Aeroespacial,\\ 
\small 12228-900 S\~ao Jos\'e dos Campos, S\~ao Paulo, Brasil \\ }

\begin{document}
\maketitle
\begin{abstract}

A recently proposed renormalization scheme can be used to deal with
nonrelativistic potential scattering exhibiting  ultraviolet divergence in
momentum space.  A numerical application of this scheme is made  in the case
of potential  scattering with $r^{-2}$ divergence for small $r$, common in
molecular and nuclear physics, by the use of cut-offs in momentum and
configuration spaces. The cut-off is finally removed in terms of a physical
observable  and model-independent result is obtained at low energies. The
expected variation of the off-shell behavior of the $t$ matrix arising from
the renormalization scheme is also discussed.

{\bf PACS Numbers 11.10.Gh, 03.80.+r., 21.30.+y}
\end{abstract}
\baselineskip .5cm
\newpage 
\section{Introduction}

Ultraviolet  divergences commonly occur in field theories, as a consequence
of point-like interactions in the original Lagrangian, both in perturbative
expansions and exact solutions.  These divergences in perturbative quantum
field theory could be eliminated by renormalization to yield a scale
\cite{books,wil} and with few exceptions the renormalized perturbative series
could not be summed.

In quantum scattering and bound states calculations, a similar situation
occurs for  potentials with certain singular behavior at short distances.  In
momentum space such problems develop ultraviolet or large momentum
divergence.  In these cases, both perturbative and exact methods fail to
obtain finite solution.

Motivated by the kind of equations that appear in quantum field theories, the
renormalization ideas have been discussed in configuration
\cite{tar1,tar2} and momentum  \cite{af,wei,afg,beg} spaces for potential
scattering, where Dirac delta, contact, or zero-range potentials were used.
Such potentials have the advantage to be simple, local and separable, with
the additional advantage that all discussion can be analytical \cite{af,afg}.

Usually, renormalization is carried out by introducing a sufficiently large
(small) cut-off in momentum (configuration)  space. With this cut-off the
divergences of the original problem are avoided and a meaningful result
obtained.  In a successful renormalization scheme the knowledge of a physical
observable at a particular energy is used to eliminate the cut-off and the
final renormalized results should be independent of the detailed
renormalization scheme in the limit when the momentum (configuration) cut-off
is removed to infinity (zero).  Renormalization of these potential models
leads to new  scales and finite physical observables  \cite{af}.

The purpose of this work is to perform numerical renormalization  of a
quantum mechanical  problem where analytical solution is not possible.  In
this way, we can study and understand better the on- and off-shell behavior
of the resulting $t$ matrix, arising from different  regularization schemes.
We consider renormalization of  $S$-wave potential scattering with  a local
attractive potential singular at $r \to 0$, given by
\begin{equation}
\label{1} V(r) = V_0 \frac{\exp{(-\mu r)}}{r^2}.
\end{equation}
In addition to performing general numerical renormalization with this
potential we shall be particularly interested in studying the expected
off-shell behavior of different regularization procedures in the
nucleon-nucleon interaction.  For a sufficiently large $|V_0|$, the
scattering problem with potential (\ref{1}) exhibits ultraviolet divergence
and does not permit solutions\cite{Landau}.  At negative energies an 
infinite number of bound states collapse to infinite binding and at positive
energies the scattering equation has noncompact kernel and does not allow
scattering solution. For a very small $|V_0|$ these theoretical problems can
be avoided but the problem becomes difficult to handle  numerically.

It is appropriate to mention that, in the limit of $\mu = 0$, the critical
strength given in Ref.\cite{Landau} is
\begin{equation}
V_{0,crit} = -\frac{1}{4}\left(\frac{\hbar^2}{2m}\right),
\label{vcrit}
\end{equation}
where $m$ is the reduced mass of the two-body system separated by distance
$r$.  For attractive (negative) potentials, with $|V_0| > |V_{0,crit}|$, 
one encounters an infinite number of bound states with infinite binding. For 
$|V_0| < |V_{0,crit}|$, these infinite number of bound states are absent. 
For a finite non-zero $\mu$, as the short
range divergent behavior of potential (\ref{1}) is unchanged, similar
properties are expected.  This behavior is apparent in our numerical results.

The motivation behind  studying this potential is that similar potentials
occur in various branches of physics, such as  molecular, nuclear and atomic
physics \cite{frank}. Also, from a mathematical point of view,
renormalization in quantum mechanics is an interesting field of interest.
Potential (\ref{1}) has particular interest in molecular and nuclear physics.
The meson-exchange field-theoretic nucleon-nucleon potentials have terms
similar to that in Eq.  (\ref{1}). Potentials with similar behavior also
appear in molecular physics.  Such terms are treated phenomenologically by
introducing an arbitrary cut-off at small $r$ or large momentum so as to cure
the original divergences and to fit some observables.  As the problem is
mathematically divergent, the results could be very sensitive to the cut-off
and it is not clear that if the momentum cut-off can be taken to infinity in
order to produce a meaningful renormalization of the problem.  In
nucleon-nucleon potentials the finite size of the hadrons sets a cut-off in
both momentum and configuration spaces, and one should work with a finite
cut-off.  However, it would be  pleasing to see that the results do not get
drastically modified as the momentum cut-off is taken to infinity.

In the present work the  above problem is renormalized numerically by the
introduction of a momentum or configuration space cut-off in different ways.
The cut-off is introduced in the Green function or the potential  in momentum
and configuration spaces, respectively. The cut-off is finally eliminated in
favor of a physical observable/information, which is chosen to be the
scattering length and the number of bound states of the system.  Eventually,
we find that by taking the cut-off to infinity, model independent results are
obtained.  An example of this scheme can  be found in Ref.\cite{am}, where 
the binding energy was used as the physical observable.  Though in the
nucleon-nucleon system one has utmost a single bound state, there could be
several bound states in atomic/molecular systems. In this study we have used
up to four bound states in order to explore the applicability of the
renormalization scheme to these general problems.

The regularization of the solution by modifying the Green function is
standard in particle and intermediate energy physics.   In atomic, molecular
and nuclear physics the traditional method is to modify the potential by form
factors. This later procedure is comparable to the modification of the
potential at short distances or large momentum as has been done in this work.
We demonstrate an equivalence between both schemes for on-shell scattering.

Though most aspects of the present study is quite general, we shall examine
some special aspects of interest to nucleon-nucleon scattering.  In addition
to the on-shell scattering under diverse situations, we also studied the
problem of  off-shell scattering with potential (\ref{1}) with a single bound
state in some details as that potential simulates nucleon-nucleon scattering.
It is well-known that the knowledge of all on- and half-on-shell $t$ matrix
elements is enough to produce the fully off-shell $t$ matrix elements
\cite{red}. This is why we only study the half-shell function \cite{gam3}.
These elements are also interesting from a physical point of view as they are
precisely the elements which can be probed experimentally in a bremsstrahlung
experiment \cite{fe}. From the present study we determine the extent of the
off-shell variation of the phenomenological nucleon-nucleon potentials
arising from different regularization schemes.

In this study, in addition to one bound state, we also considered up to four
bound states for potential (\ref{1}). The appearance of few bound states is
common in atomic and molecular physics, though in the nucleon-nucleon system
commented above there is at most one bound state. \ The successful
renormalization of potential (\ref{1}) in the presence of several bound
states suggests that the present renormalization scheme should be applicable
to similar potentials in atomic and molecular physics.

Similar renormalization has been performed recently in quantum mechanics for
the one-dimensional $x^{-2}$ potential and the three-dimensional Hulthen
potential by Gupta and Rajeev \cite{gupta}, where the interest of such
potentials in polymers has been emphasized  \cite{parisi}.

The plan of the paper is as follows. In section II,  we briefly describe the
renormalization scheme in quantum mechanics, in case of a general local
potential, following Ref. \cite{afg}.  In section III we describe the
different regularization procedures   applied to the scattering equation for
the potential given by Eq. (\ref{1}).  Finally, in section IV, we present
results and main conclusions.

\section{Renormalization in Quantum Mechanics}

In operator notation, the Lippmann-Schwinger equation  for the $t$ matrix $T$
is given by
\begin{equation}
\label{2}T(E)=V+VG_0(E)T(E),
\end{equation}
where $G_0(E)$ is the free Green function operator,
\begin{equation} 
\label{3} G_0(E)=(E-H_0+i0)^{-1},
\end{equation}
$H_0$ is the free Hamiltonian and $E$ is the center-of-mass energy.  $V$ is
the potential that, in the present case, is given by Eq. (\ref{1}).

The partial-wave projection of a matrix element of the operator $\cal O$
($\equiv T$ or $V$) in momentum space, for a spherically symmetric potential,
is given by
\begin{eqnarray}
\langle {\bf p}|{\cal O}|{\bf q}\rangle &\equiv& 
\frac{2}{\pi}
\sum_{l=0}^{\infty}{\cal O}_l(p,q)\sum_{m=-l}^l Y_{l,m}^*(\Omega_p)
Y_{l,m}(\Omega_q) \nonumber \\
&=&\frac{2}{\pi}
\sum_{l=0}^{\infty}{\cal O}_l(p,q)\frac{2l+1}{4\pi}
P_l\left(\frac{{\bf p.q}}{p q}\right),
\label{proj}
\end{eqnarray}
where $Y_{l,m}(\Omega_p)$ are the usual spherical harmonics normalized to
one, $\Omega_p$ represents the two polar angles of ${\bf p}$, and $p = |{\bf
p}|$. $P_l(x)$ is the usual Legendre polynomial of order $l$ and ${\cal
O}_l(p,q)$ is the coefficient of the expansion.

Using Eq.(\ref{proj}) in Eq.(\ref{2}) for a central potential, after the
integrations in the angular variables (where we use the orthonormalization
condition of the spherical harmonics), we obtain the corresponding partial
wave projection of Eq. (\ref{2}):
\begin{equation} 
\label{4}
T_l(p,p';E)=V_l(p,p')+\frac 2\pi \int_0^\infty {q^2dq
V_l(p,q)}G_0(q;k^2)T_l(q,p';E),
\end{equation}
where $E=k^2$, in units such that $\hbar =2m=1$, with $m$ the reduced mass.
The momentum-space free Green function is given by
\begin{equation}
G_0(q;k^2) = (k^2-q^2+ i0)^{-1}.
\end{equation}
With the definition given in Eq.(\ref{proj}), applying the Fourier transform
of the local potential $V(r)$ to momentum space, we can easily obtain the
momentum space partial wave coefficients, as detailed in the following. From
Eq.(\ref{proj}), by defining $x\equiv {\bf p.q}/(p q)$ and applying the usual
normalization of the Legendre polynomials, we have
\begin{eqnarray}
V_l(p,q) &=& \pi^2 \int_{-1}^1 dx P_l(x)\langle {\bf p}|V|{\bf q}\rangle
\nonumber \\
&=& \pi^2 \int_{-1}^1 dx P_l(x) \frac{1}{(2\pi)^3}\int d^3r  V(r) 
\exp {\left( i ({\bf p-q}){\bf .r}   \right)  } 
\nonumber \\
&=& \frac{1}{2}\int_0^\infty r^2 dr V(r) \int_{-1}^1 dx P_l(x)
\frac{\sin(|{\bf p - q}|r)}{|{\bf p - q}|r}.
\end{eqnarray}
Next, we use the addition theorem given in Eq.(10.1.45) of Ref.\cite{abram}:
\begin{equation}
\frac{\sin(|{\bf p - q}|r)}{|{\bf p - q}|r} =
j_0\left(r\sqrt{p^2+q^2-2pqx} \right) = 
\sum_{n=0}^\infty (2n+1)j_n(pr)j_n(qr)P_n(x),
\end{equation}
where $j_l(z)$ are spherical Bessel functions of order $l$ (see chapter 10 
of Ref.\cite{abram}).
Finally, using the normalization condition of the Legendre polynomials, we
obtain a simplified expression for our definition of the $l-$wave projection
of a matrix element of local potential in momentum space:
\begin{equation}
V_l(p,q) = \int_0^\infty r^2 dr V(r) j_l(pr)j_l(qr).
\label{vl}
\end{equation}
 
As we are going to consider only the $S$-wave ($l=0$) case, in the following
we simplify the notation by neglecting the $l$ indices.

With the above partial wave projection, the $S$-wave scattering length $a$
and the off-shell function $f(p,k)$ are defined, respectively,  by
\begin{equation}
a=T(0,0;0),
\end{equation}
and
\begin{equation}
f(p,k)=\frac{T(p,k;k^2)}{T(k,k;k^2)}.\label{hf}
\end{equation}

In Ref. \cite{afg}, a constant potential in momentum space, $V(p^{\prime
},p)=\lambda $, was chosen to exemplify the application of the
renormalization scheme in Quantum Mechanics. With such potential the above
equation presents ultraviolet divergence, because the kernel of the integral
equation is noncompact and Eq. (\ref{4}) does not have scattering solution.
  
So, to give meaning to Eq. (\ref{4}), when it presents ultraviolet
divergence, a new regularized equation should be defined, where the  Green
function  or the potential is replaced by a regularized one with a cut-off
parameter in momentum space $(\Lambda)$ or configuration space ($\alpha$),
respectively.  Such regularization truncates the large-momentum parts of this
equation and eliminates the ultraviolet divergence.  The cut-off parameter is
finally eliminated  in terms of a physical observable, in order to yield a
renormalized $t$ matrix which should be independent of the regularization
procedure. The Green function and potential regularizations should lead to
equivalent results, when the cut-off is removed by taking the limit $\Lambda
\to 0$ or $\alpha \to 0$, respectively.  We show in the following that the
phase shifts calculated through a Green function regularization is very
similar to those calculated through a potential regularization in the above
limit.
 
One way of introducing a regularized  Green function is to  multiply the free
Green function by a regulator function $v (q,\Lambda ;k)$ that contains a
smooth cut-off $\Lambda (>k)$, such that
\begin{equation}
G_R(q,\Lambda;k^2) 
\equiv \frac{v^2(q,\Lambda ;k^2)} {(k^2-q^2+i0)}.
\label{5}
\end{equation}
The function $v(q,\Lambda ;k)$ satisfies the conditions
\begin{equation}
v(k,\Lambda ;k) = 1 ,\label{6}\end{equation}
and
\begin{equation}
 \lim_{\Lambda \to \infty} v(q,\Lambda ;k) = 1.
\label{6a}
\end{equation}

For an appropriate  $v(q,\Lambda ;k)$ the regularized scattering equation,
\begin{equation}
T_R(p,p';E)=V(p,p')+\frac 2\pi \int_0^\infty 
{q^2dq V(p,q) G_R(q,\Lambda;k^2) T_R(q,p';E)},\label{7}
\end{equation}
has no ultraviolet divergence.

In case of a constant potential in momentum space, the following
regulator function was introduced in Ref.  \cite{afg}:
\begin{equation}
v^2(q,\Lambda ;k) \equiv \frac{(\Lambda^2+k^2)}{(\Lambda^2+q^2)}.
\label{8}
\end{equation}
For other potentials  presenting  ultraviolet divergence, one can try the
same  regulator function $v(q,\Lambda ;k)$, or, (in case of stronger
divergence) replace it by another convenient function.

The regularization procedure used above guarantees unitarity because the
imaginary part of the Green function is unaffected.  However, the limit
$\Lambda \to \infty $ reduces the regularized Green function to the free
Green function and the original ultraviolet divergence reappears.  The
renormalized $t$ matrix is obtained by eliminating the cut-off $\Lambda$ in
favor of a physical observable at some specific energy.

The above regularization through the Green function, as given in Ref.
\cite{afg}, is  on-shell equivalent to a regularization through the
potential. To demonstrate this, let us regularize Eq.  (\ref{4}) using the
following regularized potential
\begin{equation}
\label{9}V_R(p,p')=v(p, \Lambda ;k) V(p,p') v(p',\Lambda ;k).
\end{equation}
and keeping the Green function unchanged.

Now the regularized $t$-matrix elements satisfy
\begin{equation}
\label{10}T_{R}'(p,p';E)=V_R(p,p')+\frac 2\pi \int_0^\infty 
{q^2dq{V_R(p,q)}G_0(q;k^2)T_{R}'(q,p';E)}.
\end{equation}
The on-shell equivalence of both schemes can be shown by the relation 
between $T_{R}$ and $T_{R}'$, that can be obtained by replacing $V_R$,
in Eq. (\ref{10}), by Eq. (\ref{9}):
\begin{eqnarray}
&&\mathop{\hskip -1cm} 
T_{R}'(p,p';E) = v(p,\Lambda ;k) T_{R}(p,p';E) v(p',\Lambda ;k)\\
&=& v(p,\Lambda ;k) \left[ 
V(p,p')+\frac 2\pi \int_0^\infty 
{q^2dq V(p,q) \frac{v^2(q,\Lambda ;k)}{(k^2-q^2+i0)} 
T_R(q,p';E)}
\right] v(p',\Lambda ;k). \nonumber 
\label{11}
\end{eqnarray}

Considering that function  $v(p,\Lambda ;k)$ should satisfy Eq.  (\ref{6}),
all the on-shell scattering observables derived by using $T_R$ or $T_{R}'$
are identical. \ Differences between the two approaches can only be detected
when calculating off-shell observables.

\section{Models for Regularization and Renormalization}

Fredholm reduction of the Lippmann-Schwinger equation for the  $t$ matrix has
numerical advantage in dealing with real algebra  and  one such scheme was
used for numerical solution.  Auxiliary Fredholm real $\Gamma$-matrix
equation was solved numerically  \cite{gam,gam2}.  The $\Gamma$-matrix
equation avoids the fixed point singularity and has a much weaker kernel.
(For such reduced kernel techniques we refer the reader to Refs. \cite{gam3}
and references therein.) When accurate numerical results are needed, the
iterative procedure applied to the $\Gamma$ matrix integral equation has
considerable advantages, compared to the  matrix inversion methods, as shown
in Refs. \cite{gam2}.

We apply in this section the renormalization scheme presented in the previous
section to solve the scattering equation with potential (\ref{1}). \ For a
large enough $|V_0|$, this potential has a singular behavior at the origin
and admits a ground state with infinite energy, which is physically
meaningless.

To test numerically the renormalization scheme we apply four regularization
procedures for $S$-wave scattering with this potential.  In two of them we
have introduced a regulator function in the Green function as described in
the previous section.  In the other two approaches we introduce a cut-off
directly in the configuration space expression of the potential, so that a
simple regulator function in momentum space cannot be  factorized.  We
describe the four schemes (A, B, C, and D) in the following.

\noindent{\bf  Choice A}:
The $S$-wave momentum-space representation of potential (\ref{1}), after
solving Eq.(\ref{vl}) for $l=0$, with help of Eq.(3.947.2) of
Ref.\cite{grad}, is given by
\begin{equation}
V(p,q)\equiv \frac{V_0}{2pq}\left[
p \; \arctan {\frac{2\mu q}{\mu^2+p^2-q^2}} +
q \; \arctan {\frac{2\mu p}{\mu^2+q^2-p^2}}
+\frac{\mu}{2}\ln \frac{\mu^2+(p-q)^2}{\mu^2+(p+q)^2}
\right],
\label{12}
\end{equation}
with the following limiting behavior 
\begin{eqnarray}
V(0,q) &=& V(q,0) = \frac{V_0}{q} \arctan\left( \frac q\mu \right).
\label{13}
\end{eqnarray}
The regulator function in the Green function is taken to be the following
cut-off step function:
\begin{equation}
v^2(p,\Lambda_A ;k)=\Theta (\Lambda_A -p) ,
\label{14}
\end{equation}
where $\Theta (x) =1 $ for $x > 0$ and =0 for $x < 0.$

\noindent{\bf Choice B}:
The potential  $V(p,q)$ is also given by  Eq. (\ref{12}), but in this case
the regulator function in the Green function is taken  to be a smooth cut-off
function:
\begin{equation}
v^2(p,\Lambda_B ;k) \equiv \frac{(\Lambda_B^2+k^2)}{(\Lambda_B^2+p^2)}.
\label{15}
\end{equation}

\noindent{\bf Choice C}:
The regularization procedure is given in the configuration space, through a
cut-off $\alpha _C$, included in the potential, such that in the limit
$\alpha _C\to 0$ the regularized potential reduces to Eq. (\ref{1}). In this
case the potential used in calculation is given by
\begin{equation}
V_C(r)=\frac{V_0\exp (-\mu r)}{r(r+\alpha_{C} )}.
\label{17}
\end{equation}
The corresponding $S$-wave momentum-space matrix element of potential
$V_C(p,q)$ is given by
\begin{eqnarray}
V_C(p,q)&=&\frac{V_0e^{\mu \alpha_C}}{4pq}\int\limits_\mu^\infty dz
{\exp(-\alpha_C z)}
\ln \left( \frac{z^2+(p+q)^2}{z^2 +(p-q)^2}\right),
\nonumber \\
&=&\frac{V_0}{4\alpha_C pq}\ln \left( \frac{\mu^2+(p+q)^2}
{\mu^2 +(p-q)^2}\right)
 - \frac{2V_0}{\alpha_C}e^{\mu \alpha_C}\int\limits_\mu^\infty dz
\frac{z\;\exp(-\alpha_C z)}{[z^2+(p+q)^2] [z^2+(p-q)^2]},\nonumber \\
&=&\frac{V_0\mu e^{\mu \alpha_C}}{4pq}\int\limits_0^1dy\frac{\exp
(-\mu \alpha
_C/y)}{y^2}\ln \left( \frac{\mu ^2+(p+q)^2y^2}{\mu ^2+(p-q)^2y^2}\right),
\label{18}
\end{eqnarray}
with the limiting behavior
\begin{equation}
V_C(p,0) = V_C(0,p) =  
V_0 e^{\mu \alpha_C}\int\limits_\mu^\infty dz\,
\frac{\exp (-\alpha _C z)}{z^2+p^2} =
V_0\mu e^{\mu \alpha_C}\int\limits_0^1dy\,
\frac{\exp (-\alpha _C\mu /y)}{\mu ^2+(p y)^2}.
\label{19}
\end{equation}

\noindent{\bf  Choice D}:
The procedure for regularization is given in configuration  space, through a
cut-off $\alpha _D$, such that in the limit $\alpha _D\to 0$, $V$ reduces to
Eq. (\ref{1}). The regularized potential for this choice is
\begin{equation}
V_D(r)=\frac{V_0\exp (-\mu r)}{(r+\alpha _D)^2},
\label {20}
\end{equation}
and the corresponding $S$-wave momentum-space potential $V_D(p,q)$ is given
by
\begin{eqnarray}
V_D(p,q)&=&
2V_0e^{\mu \alpha _D}\int\limits_\mu^\infty dz \frac{z(z-\mu)\exp
(-\alpha_D z)}{[z^2+(p+q)^2][z^2+(p-q)^2]}
\nonumber \\
&=&2V_0e^{\mu \alpha _D}\mu ^3\int\limits_0^1dy\frac{(1-y)\exp (-\mu
\alpha _D/y)}{\mu ^4+2\mu ^2y^2(p^2+q^2)+(p^2-q^2)^2y^4}.
\label{21}
\end{eqnarray}

For  choices A and B the regulator function  is introduced in the Green
function, as explained in section II. For  choices C and D, the cut-off is
introduced directly in the configuration space expression of the potential,
before the Fourier transform is taken, such that a regulator function cannot
be simply factorized in momentum space, as in choices A and B.

The results for the various renormalization schemes is expected to be
independent of the detailed regularization procedures. Such expectation is
verified in numerical calculation in the next section, and its limitation
discussed.

\section{Numerical Results and Conclusions}

In the following we present  the results for phase-shifts, obtained by using
four different regularization procedures  of the previous section.  The
reduced mass $m$  has been taken to be appropriate for the nucleon-nucleon
system; $\hbar^2/(2m)$ = 41.47 MeV fm$^2$. The range parameter $\mu$ of the
potential has been taken to be 1 fm$^{-1}$.  Though we have chosen nuclear
units in our calculation, most aspects of this numerical study are of general
interest.

In the renormalization scheme we define the cut-off parameter through one of
the  choices of regularization (for example, choice A), given by Eq.
(\ref{14}), and fix the scattering length to the experimental  value ($a=5.4
$ fm, for the $S$-wave spin triplet nucleon-nucleon scattering) by adjusting
the strength $V_0$ of the potential. The scattering length does not fix the
number of bound states of the system and  the strength of the potential fixes
this number.  In present renormalization we considered up to four bound
states of the system. For  other three regularization procedures (choices B,
C and D) the strength $V_0$ and the corresponding number of bound states
(obtained with  choice A) are kept fixed together with the scattering length
by redefining the corresponding cut-off parameters. \ With this procedure we
guarantee that we are dealing with the renormalization of the same  physical
interaction via four regularization procedures. Presently, the physical
observables that we fix by renormalization are the scattering length  and the
number of bound states.  The numerical parameters used in calculation are
given in Table 1.

Though we consider up to four bound states of potential (\ref{1}) and also
consider the limit of very large momentum space cut-off, the situation of
interest in the nucleon-nucleon system is the case of one bound state, the
deuteron, and a finite cut-off determined by the ``size" of the problem.
Hence, for the nucleon-nucleon system the first row in Table 1 is of
interest, where the momentum space cut-off is set at 10 fm$^{-1}$. We shall
explore the off-shell behavior of the nucleon-nucleon scattering under this
situation.

In each of  Figs. 1-3, we display four sets of curves for phase shift versus
energy.  In order to have compact figures, in all case we set the zero-energy
phase shift at $\pi$.  Each set is indicated by an ellipse and  labeled by
the number of bound states. \ In each set, there are four curves that
represent  four different regularization procedures.  Schemes (A$-$D) are
conveniently indicated inside the figures.    The cut-off parameters of the
four schemes are $\Lambda_A$, $\Lambda_B$, $\alpha_C$, and $\alpha_D$.  In
the renormalization limit the momentum cut-offs $\Lambda_A$ and $\Lambda_B$
should be large, and the configuration cut-offs $\alpha_C$ and $\alpha_D$
should be small. Also, the cut-offs should be large compared to the on-shell
momentum. In schemes A $-$ D this means $\Lambda_A^2 >> k^2$, $\Lambda_B^2 >>
k^2$, $\alpha_C^{-2} >> k^2$, and $\alpha_D^{-2} >> k^2$, respectively.

By comparing the curves in each set of Fig. 1, we observe that the phase
shifts are numerically very close for energies up to 50 MeV, that correspond
to $k^2 \sim 1$ fm$^{-2}$, or 0.01 $\Lambda_A ^2$.  The deviations start to
increase as  energies approach the cut-off limit.

In Fig. 2, we repeat the same plots as  in Fig. 1 with larger momentum
cut-offs, $\Lambda_A$  and $\Lambda_B$, or reduced configuration cut-offs,
$\alpha_C$ and $\alpha_D$. By increasing the momentum cut-off we approximate
the renormalization limit of large  cut-off.  \ In this case, comparing the
curves in each set, we observe that the phase-shifts are numerically very
close for energies up to 500 MeV, that correspond to $k^2
\sim$ 10 fm$^{-2}$ $\approx$ 0.025 $\Lambda_A ^2$.

The plots of  Figs. 1 and 2 are repeated in Fig. 3 for a further increased
momentum cut-off (reduced configuration cut-off).
\ As we can see, the phase-shifts, for each set of plots
that have the same number of bound states, are very close to each other in
all extension of the energies shown in the graph. The energies can go close
to 1000 MeV, or $k^2 \sim $ 20 fm$^{-2} \approx  0.025 \Lambda_A^2$.

As we can see from the three figures, the limit of validity of
renormalization  increases fast with the increase of the momentum cut-off. As
expected, the renormalization  become independent of different regularization
procedures, if the momentum cut-off is large enough compared to the on shell
momentum or equivalently, the configuration cut-off is small enough.  In this
limit the domain of renormalization extends roughly up to a c.m.  energy $
E\approx   0.025 E_\Lambda$  where $E_\Lambda=\hbar^2\Lambda^2/(2m)$.

For $k << \mu$, pure $S$-wave prevails and the scattering phase shift
decreases linearly as $k$: $\delta \approx \pi -ka$.  This behaviour is
consistent with the phase shift of the contact interaction \cite{af}: $\tan
(\delta) = -ka$, which for small $k$ reduces to $\delta \approx \pi - ka$.
This is clearly represented in Figs. 1$-$3. This behavior does not depend on
the details of the potentials and on the number of bound states.

From Table 1, and Figs. 1$-$3, it is interesting to observe that, even for
$\mu \ne 0$, in the case of large cut-offs, the critical strength to produce
four bound states is consistent with the limit given in Eq.(\ref{vcrit}).
For the nucleon-nucleon system $V_{0,crit} = -$ 10.3675 MeV fm$^2$.  Note
that in Fig.3 we have a cut-off with 30 fm$^{-1}$ and the corresponding $V_0$
($= -$12.482 MeV.fm$^2$).  In the limit of very large cut-off and large
number of bound states, $V_0$ should approach its critical value.

The cut-off behavior of the renormalization scheme  is shown in Fig. 4. In
this figure we plot the dispersion in phase shifts among the four schemes A
$-$ D versus the cut-off parameter $\Lambda_A$ in the case with  four bound
states.  To show the dispersion  in phase shifts we plot  the differences
between the phase shifts $\Delta \delta$ that show maximum deviation at a
fixed energy:  choices A and B. The three curves correspond to dispersion at
three energies 150 MeV, 500 MeV, and 1000 MeV.  As we can observe in this
figure, the dispersion  approaches  zero as we go to infinite momentum
cut-off limit, and the results become independent of the regularization
procedure that have been used.

We have studied systems that can have up to 4 bound states, but as pointed
out before, the situation of interest in the nucleon-nucleon system is the
first row in Table 1 where there is only one bound state and a finite
momentum space cut-off of about 10 fm$^{-1}$. The situations with more than
one bound states are of interest in atomic/molecular physics.  Of the three
momentum  cut-offs used in this work the one at 10 fm$^{-1}$ seems to be more
appropriate for the nucleon-nucleon system.

The present work can be used to conclude about the expected off-shell
variation of different nucleon-nucleon potential models with the same
physical content.  Here we have used one constraint on the regularization
scheme so as to fit the scattering length.  In realistic situations several
constraints are used in phenomenological nucleon-nucleon potentials so that
the on-shell phase shifts of the different models are practically the same.
However, $t$ matrices generated from such potentials may have large
off-shell variations as pointed out in Refs.
\cite{red,fe}.  As fully off-shell $t$ matrices are determined in terms of
the half-shell $t$ matrices \cite{red} we plot in Fig.~5 the half-shell
function (\ref{hf}) for the various schemes at following energies: $E_{c.m.}$
= 0, 37.5 MeV, 140 MeV, and 300 MeV. These were the energies explored by
Fearing in his study of bremsstrahlung \cite{fe} with different realistic
potentials.

The half-shell variation in Fig. 5 is not a consequence of the on-shell
variations observed in Figs. 1$-$3. At zero energy the four on-shell $t$
matrices are the same but they may have distinct off-shell behaviors.  In the
two cases (choices A and B) where momentum cut-offs were employed the
regularization schemes are similar and there is no off-shell variation at low
energies. The same is true for choices C and D. For the parameters used in
the calculation we note that the momentum space cut-off functions (\ref{14})
and (\ref{15}) of choices A and B are quite different numerically.  The same
off-shell behavior generated in these two choices suggest that such momentum
space cut-offs in two phenomenological models should not lead to large
off-shell variations independent of the values of the parameters used.

The off-shell variation of Fig.~5 is significantly lower than that observed
by Fearing in different theoretical models. The larger half-shell variation
observed by him was possibly due to differences in the physical contents of
the phenomenological nucleon-nucleon potential models. This indicates
different intermediate-range (2 $-$ 4 fm)  behaviors of those realistic
potentials. In the present study the long- and intermediate-range behaviors
of the models are  kept the same, while the short-range part of these
potentials are varied in the different regularization schemes.    This
variation should increase as the momentum space cut-off $\Lambda$ is reduced.
For a fixed $\Lambda$ it should increase as the momentum variables $k$ or $p$
of the half-shell function are increased. The  present half-shell variations
noted in Fig. 5 are arising from the inherent differences in different
regularization schemes.  Similar half-shell variation(s) are expected in
different phenomenological nucleon-nucleon potentials with same physical
content.

In summary, we have tested numerically renormalization schemes using a
singular local potential, that diverges at $r \to 0$ as the inverse of $r^2$.
The scattering integral equation  exhibits renormalizable ultraviolet
divergences in this case. We performed numerical renormalizations with four
regularization procedures, two with a cut-off in momentum space and two with
a cut-off in configuration space.  We find  that the renormalization
techniques are applicable to this case, such that, for lower energies
compared to the cut-off, the  results are independent of regularization
procedures used to regularize the original equation.  In the case of the
nucleon-nucleon system the regularization scheme should maintain a single
bound state and employ a finite cut-off. In that case we studied the
off-shell variation of the different regularization schemes and found that
the different phenomenological nucleon-nucleon potential models have larger
off-shell variations \cite{fe}  than found in the present study. Different
intermediate-range behaviors of the usual phenomenological potentials are
supposed to be responsible for larger off-shell variations.  The present
renormalization in the presence of up to four bound states is of relevance in
atomic and molecular physics, and suggest that similar potentials in atomic
and molecular physics can also be successfully renormalized in the present
scheme.

We thank the Conselho Nacional de Desenvol\-vimento Cient\'\i fico e
Tecnol\'ogico, Funda\-\c c\~ao de Amparo \`a Pesquisa do Estado de S\~ao
Paulo, Coor\-dena\c c\~ao de Aper\-fei\c coa\-mento de Pessoal do N\'\i vel
Superior and Finan\-ciadora de Estu\-dos e Projetos of Brazil for partial
financial support.

\newpage
\begin{table}
\renewcommand{\baselinestretch}{1.5}
\caption{Numerical parameters used in the four regularization schemes.
$\Lambda_A$ is the regulator used in choice A, $\Lambda_B$ the corresponding
regulator of choice B, $\alpha_{C}$ and $\alpha_{D}$ are the regulators used
in the choices C and D  respectively.  $N$ is the number of bound states.
For each set of data with the same regulator $\Lambda_A$, there is a
corresponding figure that is indicated in the first column in the left. The
scattering length is kept fixed at 5.4 fm in all cases, independent of the
number of bound states.}

\begin{center}
\begin{tabular}{|ccccccc|}\hline \hline
Fig.  & N &$V_0$ (MeV fm$^2$) &$\Lambda_A $(fm$^{-1}$) &$\Lambda_B$ (fm$^{-1}$)
  &$\alpha_C$ (fm) &$\alpha_D $(fm) \\
\hline \hline
& &  &  &  &  & \\ 
&1& $-$0.6348 & 10 & 12.145 & 0.020984 & 0.008942 \\ 
1 & 2& $-$3.6000 & 10 & 8.830  & 0.131200 & 0.044080 \\ 
& 3& $-$14.030 & 10 & 6.488  & 0.403600 & 0.112400 \\ 
& 4& $-$40.132 & 10 & 5.200  & 0.870300 & 0.211380 \\ 
& &  &  &  &  & \\ 
& 1& $-$0.5306 & 20 & 25.420 & 0.008515 & 0.003693 \\ 
2& 2& $-$2.2325 & 20 & 19.896 & 0.042190 & 0.015260 \\ 
& 3& $-$7.2221 & 20 & 15.395 & 0.119400 & 0.036880 \\ 
& 4& $-$17.705 & 20 & 12.838 & 0.238200 & 0.065960 \\ 
& &  &  &  &  & \\ 
& 1& $-$0.48935 & 30 & 38.930 & 0.005158 & 0.002261 \\ 
3& 2& $-$1.80350 & 30 & 31.280 & 0.022850 & 0.008420 \\ 
& 3& $-$5.39250 & 30 & 24.785 & 0.062700 & 0.020185 \\ 
& 4& $-$12.4820 & 30 & 20.917 & 0.122570 & 0.035700\\
& & &  & &  & \\
\hline  \hline
 \end{tabular}
 \end{center}
\end{table}
\vfill \eject

{\bf Figure Caption}\\
\begin{description}
\item[Fig. 1] 
Renormalized phase shifts versus energy using four regularization procedures
described in the text. The regulator used for choice A is $\Lambda_A
=10\;$fm$^{-1}$. The corresponding regulators for the other three choices,
and the corresponding strength $V_0$ that maintains the scattering length
fixed at 5.4 fm, when varying the number of bound states $N$, are given in
the first set of Table 1.
\item[Fig. 2] 
The same as in Fig. 1, using the parameters given in the second set of Table
1, that have $\Lambda_A = 20\;$fm$^{-1}$.
\item[Fig. 3] 
The same as in Fig. 1, using the parameters given in the third set of Table
1, that have  $\Lambda_A = 30\;$fm$^{-1}$.
\item[Fig. 4]
The dispersion in phase shifts $\Delta \delta$ versus the momentum cut-off
$\Lambda_A$, for the energies 150 MeV, 500 MeV, and 1000 MeV, in the case we
have four bound states. We define the dispersion by the difference between
the phase shifts of choices A and B, which are enhanced for larger number of
bound states $N$, as shown in Figs. 1-3.
\item[Fig. 5]
Off-shell function $f(p,k)$ versus $p$, for different regularization schemes
corresponding to the first row of Table 1.  The curves are labeled by the
respective center of mass energies, in units of MeV.
\end{description}
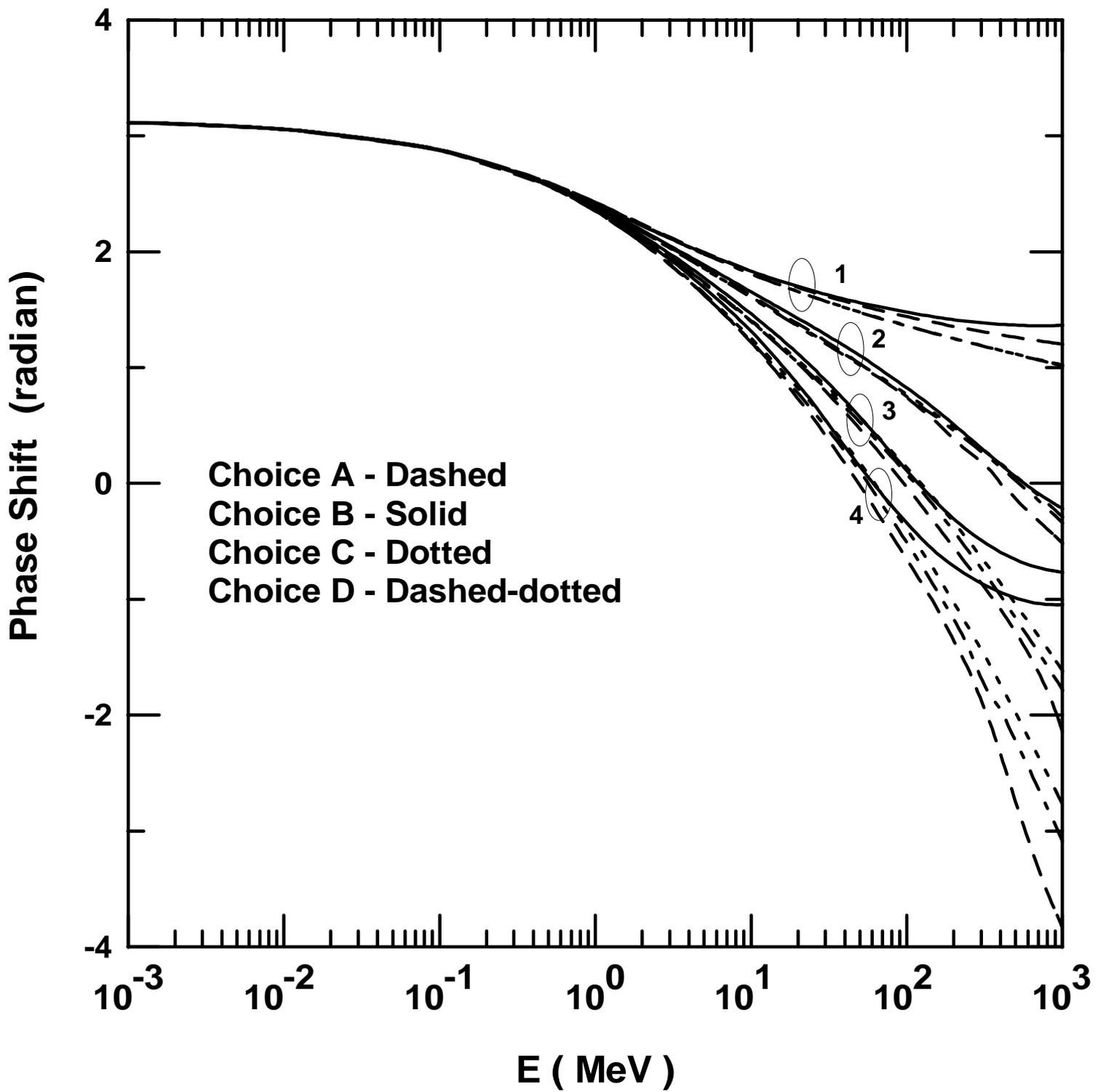
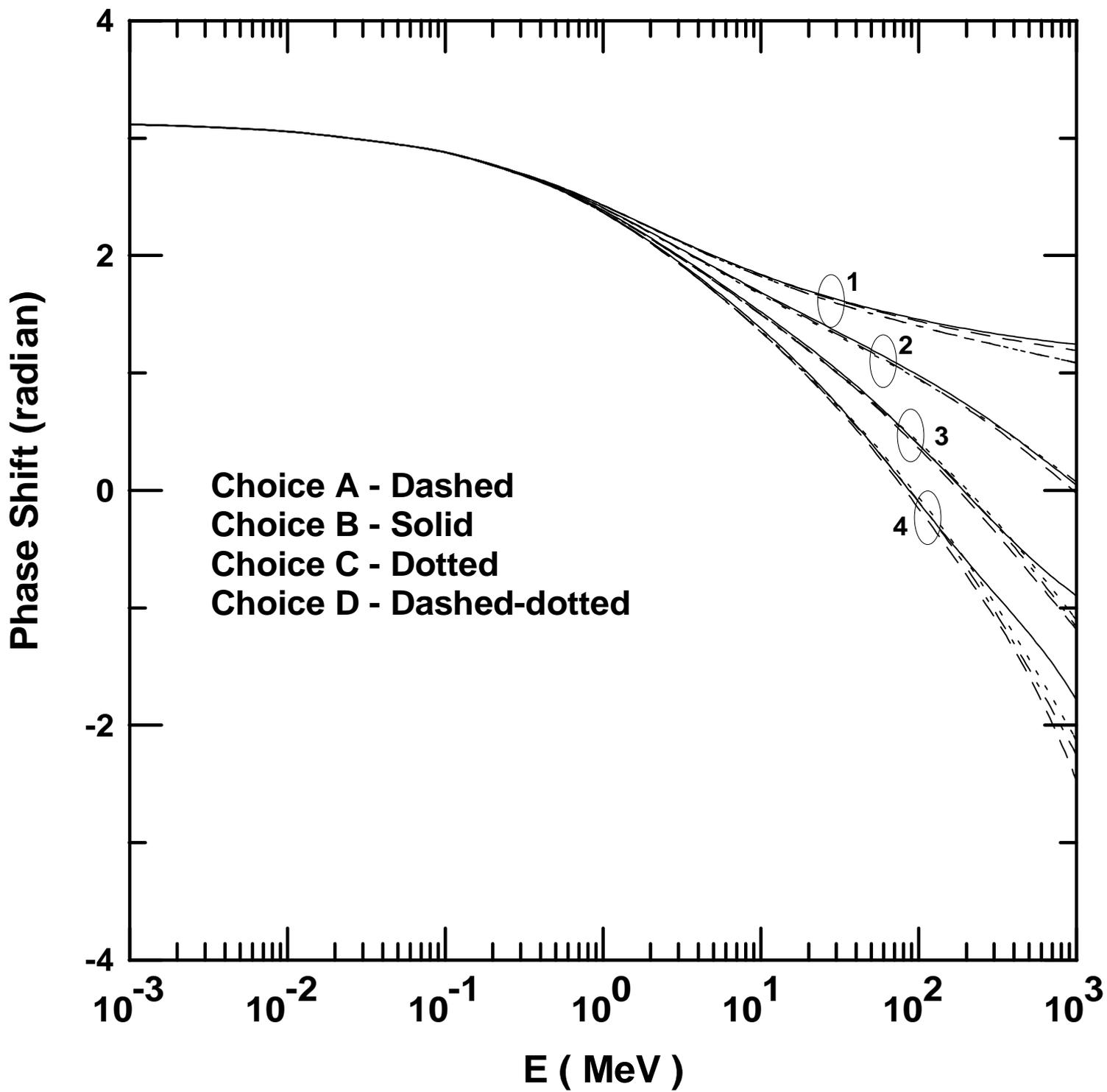
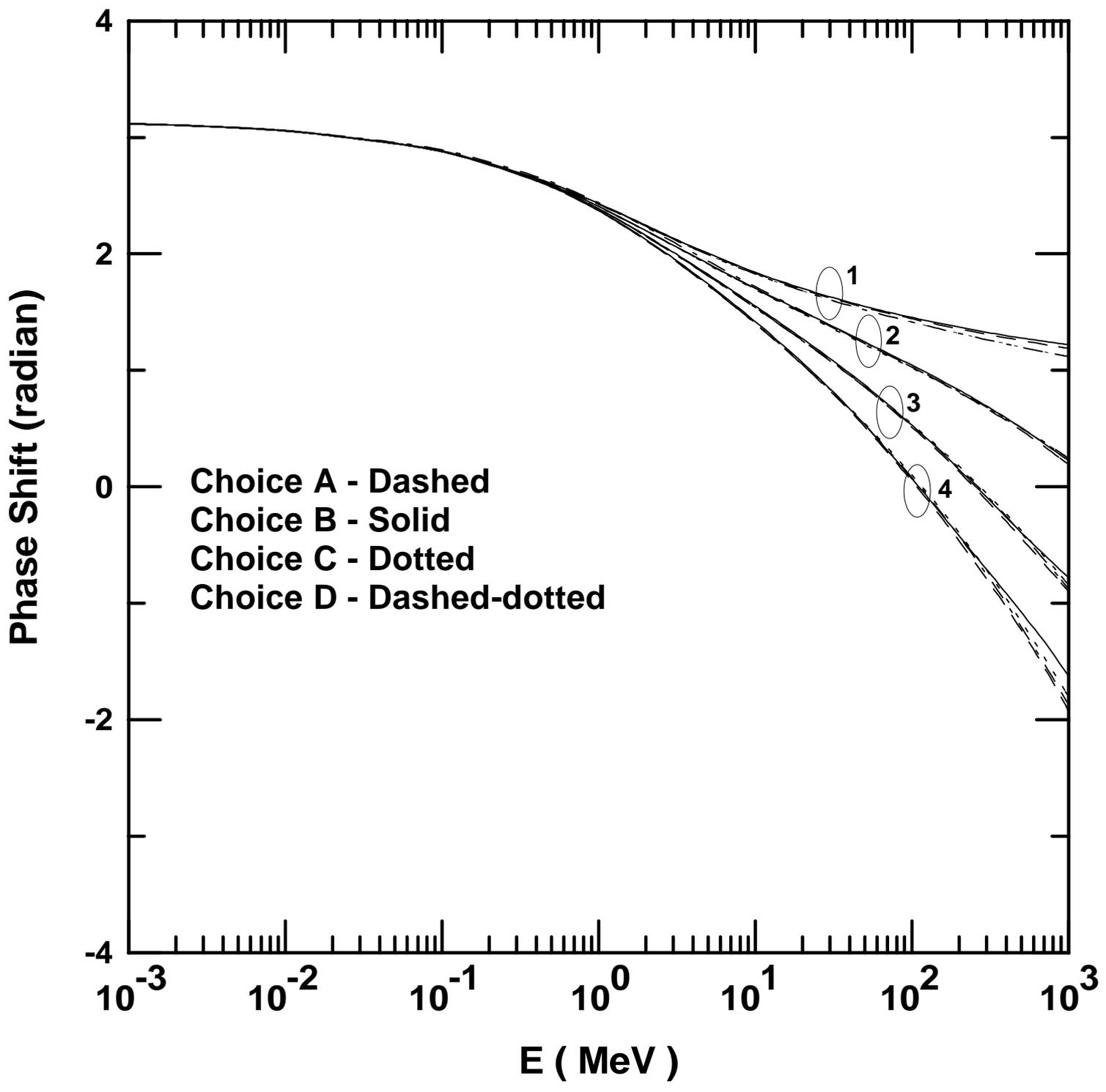
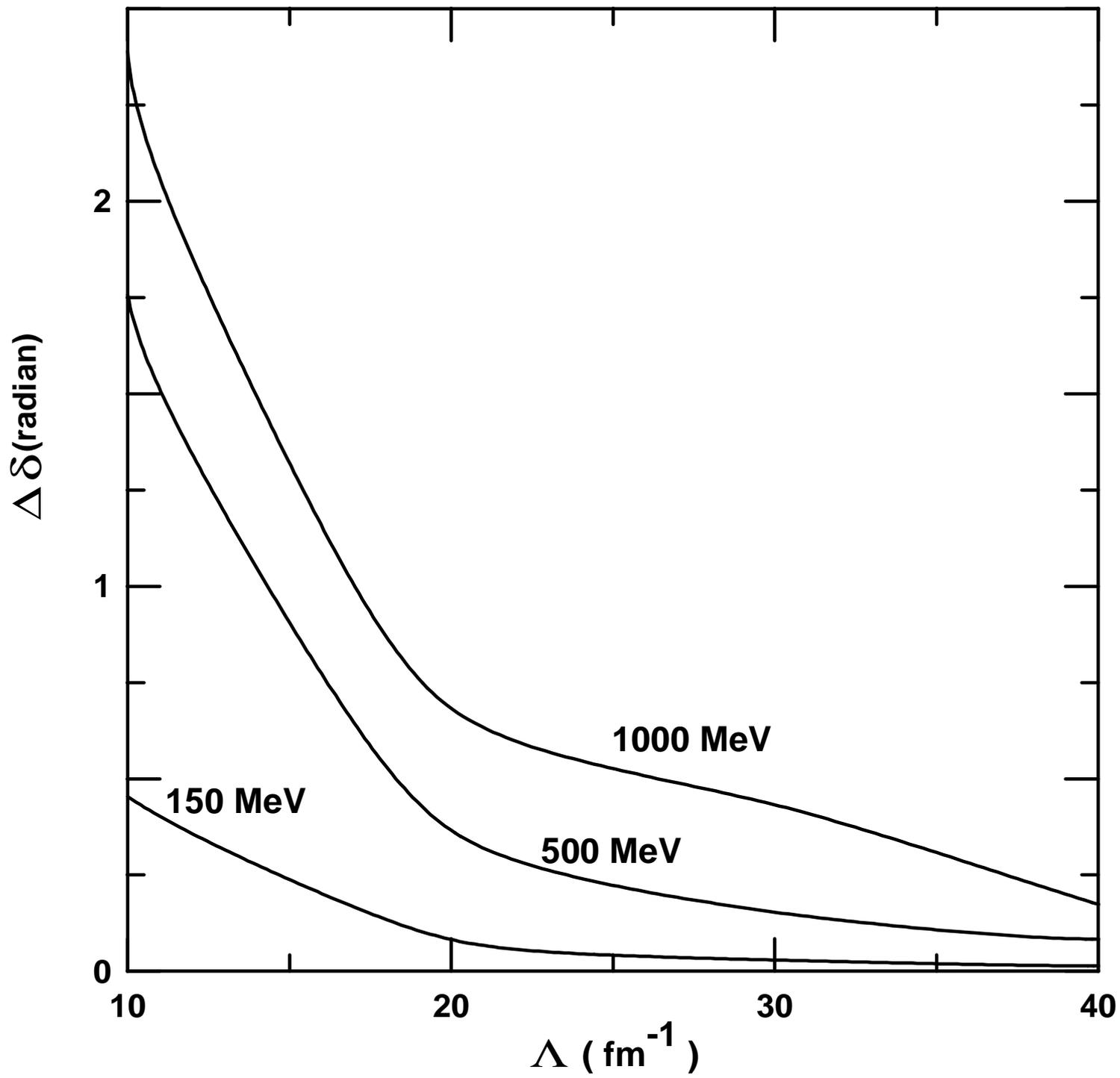
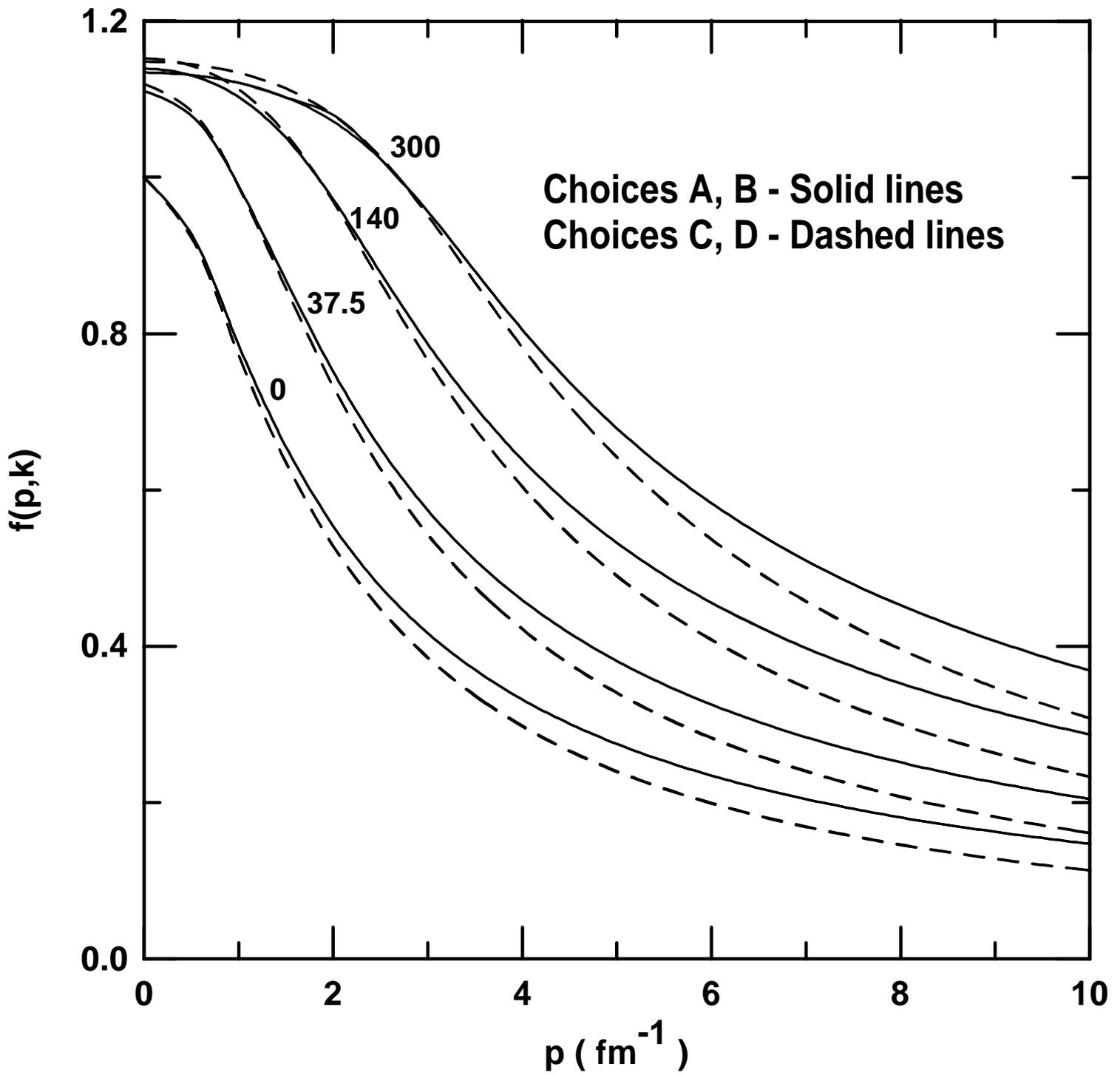
\end{document}